\documentclass[12pt,a4paper]{article}
\usepackage{amsmath}
\usepackage[dvips]{graphicx}
\input epsf
\usepackage{times}
\begin{document}
\begin{titlepage}
\title{Regularities of
 vector--meson electroproduction: transitory effects or early asymptotics?}
\author{ S. M. Troshin, N. E. Tyurin\\[1ex]
\small  \it  Institute for High Energy Physics,\\
\small  \it Protvino, Moscow Region, 142281, Russia}
\normalsize
\date{}
\maketitle

\begin{abstract}
We discuss recent HERA results on vector--meson electroproduction\\
$\gamma^*p\to Vp$ and demonstrate that universality of the
initial and final state interactions responsible for the transition between the
on-- and off--mass shell states allows
to explain energy independence of the ratio   of exclusive
$\rho$ electroproduction cross section to the total cross section. We also predict
explicit mass dependence of this ratio for other vector mesons.
\end{abstract}
\end{titlepage}
\setcounter{page}{2}

\section*{Introduction}
Besides  the  studies of the structure function $F_2$ in DIS at low $x$
the important measurements of the cross--sections of  elastic
vector meson production were performed in the experiments H1 and
ZEUS at HERA \cite{zosa,melld}.

As it follows from this data  the
 integral  cross section
of the elastic vector meson production $\sigma^{V}_{\gamma^* p}(W^2,Q^2)$
 increases with energy in a way
similar to the
$\sigma^{tot}_{\gamma^*p}(W^2, Q^2)$ dependence
 on  $W^2$ \cite{her}.
It appeared also that the growth of  the vector--meson
electroproduction cross--section  with energy is steeper for
heavier vector mesons. Similar effect takes also place when the virtuality
$Q^2$ increases.

The most recent  data of ZEUS Collaboration evidently demonstrated an
 energy independence of the
 ratio of the cross section of exclusive
$\rho$ electroproduction to the total cross section \cite{levy}.
Such behaviour of
this ratio is at variance with perturbative QCD results \cite{brod}, Regge and dipole
approaches \cite{levy,bial}. Recent review of the related   problems and successes
of the various theoretical approaches can be
found in \cite{land}.

Of course, the energy range of the available data is limited and
the above mentioned contradiction could probably be avoided due to fine tuning of
 the appropriate models. But this is merely  a general statement and it needs to be checked
 in each particular case.
Meanwhile, it would be interesting to pay an attention to a model where
such  an energy independence is an inherent one. To be more specific
this energy independence was  obtained in the approach
based on the off-shell extension of the $s$--channel unitarity \cite{epj02}.
It is worth to note also that the approach developed in \cite{epj02} and its application
to the elastic vector meson production  processes
$\gamma^*p\to Vp$ leads to mass dependence of the corresponding
cross--sections  which is in a
good agreement with the experiment \cite{perf} and it allows us to make  predictions for the
explicit mass dependence of the ratio
 $r_V=\sigma^{V}_{\gamma^* p}(W^2,Q^2)/\sigma^{tot}_{\gamma^* p}(W^2,Q^2)$.

\section*{Vector--meson electroproduction}

There is no  universal, generally accepted
method to obey unitarity of the scattering matrix.
However, long time ago  arguments based on analytical properties
 of the scattering
amplitude  were put forward \cite{blan} in favor of the rational form of unitarization.
Unitarity relations can be written for both
real and virtual external particles scattering amplitudes.
However, implications of unitarity  are different
 for the scattering of real and virtual
particles.
The extension of the $U$--matrix unitarization scheme
 (rational form of unitarization)
for the off-shell
scattering was considered in \cite{epj02}. It was supposed as usual that
the virtual  photon fluctuates into a quark--antiquark
pair $q \bar q$ and this pair can
be treated as an effective virtual  vector meson state.
There were considered limitations the unitarity provides for the $\gamma^* p$--total
 cross-sections and geometrical effects in the
 energy dependence of $\sigma^{tot}_{\gamma^* p}$.
It was shown that the solution of the extended unitarity equations augmented
by an assumption of
 the $Q^2$--dependent constituent quark
 interaction radius   results in the following
dependence at high energies:
\begin{equation}\label{sigt}
\sigma^{tot}_{\gamma^* p}\sim (W^2)^{\lambda(Q^2)},
\end{equation}
where
 $\lambda(Q^2)$ is saturated
 at large values of $Q^2$ and reaches unity. However, off--shell unitarity
 does not require transformation of this power-like dependence into a logarithmic
 one at asymptotical energies. Thus,  power--like behaviour
 of the cross--sections
with exponent dependent on virtuality
 could be of an asymptotical nature and have a  physical ground.
  It should not  be regarded merely as a transitory behaviour
or a convenient way to represent the data.

The extended unitarity  for the off--mass--shell amplitudes $F^{**}$ and
$F^*$ has a structure  similar to the equation for the on--shell
amplitude $F$ but in the former case it relates the
different amplitudes\footnote{We denoted that way the
amplitudes when both initial and final mesons  are off mass
shell, only initial meson is off mass shell and both mesons are on
mass shell, respectively. Note that $\sigma^{tot}_{\gamma^* p}$
 is determined by the imaginary part of the amplitude $F^{**}$,
  whereas $\sigma^{V}_{\gamma^* p}$ is
 determined by the square of  another amplitude $F^*$.}.
The important point in the solution of the extended unitarity
is the factorization in the impact parameter representation
at the level of the input dynamical
quantity~---~$U$--matrix:
\begin{equation}
U^{**}(s,b,Q^2)U(s,b)-[U^{*}(s,b,Q^2)]^2=0.\label{zr}
\end{equation}
Eq. (\ref{zr}) reflects universality of the initial and final state interactions
when transition between on-- and off--mass shell states occurs.
Despite that such factorization does not
survive  at the level of the
amplitudes $F^{**}(s,t,Q^2)$, $F^*(s,t,Q^2)$
and $F(s,t)$ (i.e.  after unitarity equations are solved and Fourier-Bessel
  transform is performed), it is essential for the energy independence of the
 ratio  of the exclusive
$\rho$ electroproduction cross section to the total cross section.

The above result (\ref{sigt}) is valid  when the interaction radius of the virtual
constituent quark is rising with virtuality $Q^2$.
The dependence of the interaction radius
\begin{equation}\label{rqvi}
r_{Q^*}=\xi(Q^2)/m_Q.
\end{equation}
on  $Q^2$ comes through the dependence
of the factor $\xi(Q^2)$ (in the on-shell limit $\xi(Q^2)\to\xi$).
The origin of the rising
 interaction radius of the virtual
constituent quark with virtuality $Q^2$
 might be of a dynamical nature and it would steam from
 the emission of the additional $q\bar q$--pairs in the
 nonperturbative  structure of a constituent quark.
Available experimental data
 are consistent with the $\ln Q^2$--dependence of the radius
 \cite{epj02}:
\[
\xi(Q^2)=\xi + a\ln\left(1+ \frac{Q^2}{Q_0^2}\right).
\]

The introduction of the $Q^2$ dependent   interaction radius of a constituent
quark, which in this approach consists of a current quark
surrounded by the  cloud of quark--antiquark pairs of different
flavors \cite{csn}, is the main issue of the off--shell extension of the
model, which provides at large values of
$W^2$
\begin{equation}\label{totv}
\sigma^{tot}_{\gamma^* p}(W^2,Q^2)\propto G(Q^2)\left(\frac{W^2}{m_Q^2}
\right)^{\lambda (Q^2)}
\ln \frac{W^2}{m_Q^2},
\end{equation}
where
\begin{equation}\label{lamb}
\lambda(Q^2)=\frac{\xi(Q^2)-\xi}{\xi(Q^2)}.
\end{equation}
 The value of parameter $\xi$
 in the model is determined by the slope of the differential cross--section
of elastic scattering at large $t$ region \cite{lang}
and it follows from the $pp$-experimental data  that $\xi=2$.

Inclusion of heavy vector meson production into this
scheme is straightforward: the virtual photon fluctuates before
the interaction with proton into the heavy quark--antiquark pair
 which constitutes
the virtual heavy vector meson state. After interaction with a proton
this state turns out into a real  heavy vector meson.

Integral exclusive (elastic) cross--section of vector meson production in
the process $\gamma^*p\to Vp$ when the vector meson in the final
state contains not necessarily  light quarks can be calculated directly:
\begin{equation}\label{elvec}
\sigma^{V}_{\gamma^* p}(W^2,Q^2)\propto G_{V}(Q^2)\left(\frac{W^2}
{{m_Q}^2}
\right)^{\lambda_{V} (Q^2)}
\ln \frac{W^2}{{m_Q}^2},
\end{equation}
where
\begin{equation}\label{lavm}
\lambda_{V}(Q^2)= \lambda (Q^2)\frac{\tilde{m}_Q}{\langle m_Q \rangle}.
\end{equation}
In Eq. (\ref{lavm}) $\tilde{m}_Q$ denotes the mass of the constituent
quarks from
the vector meson and $\langle m_Q \rangle$ is the mean constituent
quark mass
of the vector meson and proton system.
Of course, for the on--shell scattering ($Q^2=0$) we have a standard Froissart--like
asymptotic energy dependence.

It is evident from Eqs. (\ref{totv}) and (\ref{elvec}) that
$\lambda_{V}(Q^2)=\lambda(Q^2)$
for the light vector mesons, i.e. the ratio
\begin{equation}\label{rv}
r_V=\sigma^{V}_{\gamma^* p}(W^2,Q^2)/\sigma^{tot}_{\gamma^* p}(W^2,Q^2)
\end{equation}
does not depend on energy for $V=\rho,\omega$.
For the case of the heavy vector meson production $J/\Psi$ and $\Upsilon $ the
 respective cross--section
rises about two times  faster than the total  cross--section;
Eq. (\ref{lavm}) results in
\[
\lambda_{J/\Psi}(Q^2)\simeq 2\lambda(Q^2),\quad
\lambda_{\Upsilon}(Q^2)\simeq 2.2\lambda(Q^2),
\]
i.e.
\[
r_{J/\Psi} \propto (W^2)^{\lambda(Q^2)},\quad
r_{\Upsilon } \propto (W^2)^{1.2\lambda(Q^2)}.
\]
Corresponding relations for the $\varphi$--meson production
 are the following
\[
\lambda_{\varphi}(Q^2)\simeq 1.3\lambda(Q^2),\quad
r_{\varphi} \propto (W^2)^{0.3\lambda(Q^2)}.
\]
In the limiting case when the vector meson
is very heavy, i.e. $\tilde m_Q\gg m_Q$ the relation
between exponents is
\[
\lambda_{V}(Q^2)=\frac{5}{2}\lambda(Q^2).
\]

The quantitative agreement  of Eq. (\ref{elvec})  with experiment
was demonstrated in \cite{perf}.
This  agreement is in favor of relation (\ref{lavm}) which provides explicit
mass dependence of the exponent  in the power--like energy dependence of cross--sections.
Thus, the power-like parameterization of the
ratio $r_V$
\[
r_V\sim (W^2)^{\lambda (Q^2){(\tilde{m}_Q-\langle m_Q \rangle)}/
{\langle m_Q \rangle}}
\]
 with $m_Q$ and $Q^2$--dependent
exponent could also have a physical ground.  It would be interested therefore
to check experimentally the predicted energy dependence of the
ratio $r_V$.

The dependence of the constituent quark
interaction radius  $r_{Q^*}=\xi(Q^2)/m_Q$ on its mass and virtuality
gets an experimental support and the
 non--universal energy asymptotical dependence Eqs. (\ref{elvec}) and (\ref{lavm}) and
predicted  in \cite{epj02}
 does not  contradict to
the high--energy experimental data on elastic
vector--meson electroproduction. Of course, as it was already mentioned in the
Introduction, the limited energy range of the available experimental data  allows
other parameterizations, e. g.  universal asymptotical Regge--type behavior
with $Q^2$--independent trajectories (cf. \cite{petr,mart,fior}), to treat
the observed experimental
 regularities as transitory effects.

It seems, however, that the scattering of virtual
particles reaches the  asymptotics much faster than the scattering of
the real particles and the $Q^2$--dependent exponent $\lambda(Q^2)$ reflects asymptotical
dependence and not the "effective" preasymptotical one.
Despite that the relation  between $\xi(Q^2)$ and $\lambda(Q^2)$
 implies  a saturation of the $Q^2$-dependence of $\lambda(Q^2)$ at large
values of $Q^2$, the power--like energy dependence itself will survive
 at asymptotical energy values.
The early asymptotics of virtual particle scattering
is correlated with the peripheral  impact parameter
 behavior of the scattering amplitude
for the  virtual particles. The respective profiles of the amplitudes $F^{**}$
and $F^{*}$ are peripheral when $\xi(Q^2)$
increases with $Q^2$ \cite{epj02}.

The energy independence of the ratio
 $r_\rho=\sigma^{\rho}_{\gamma^* p}/\sigma^{tot}_{\gamma^* p}$
  reflects
 universality of the
initial and final state interactions responsible for the transition between the
on-- and off--mass shell states.
This universality is a quite  natural assumption
leading to factorization Eq. (\ref{zr}) at the level of the $U$--matrix \cite{epj02}.
Thus, the  off--shell unitarity is the principal origin of the
observed energy independence of the ratio $r_\rho$.
\section*{Acnowledgement}
We are grateful to V.~Petrov and
 E.~Martynov  for  the interesting
 discussions.

\end{document}